# Inter-brain substrates of role switching during mother-child interaction

Yamin Li[1,5,#] Saishuang Wu[2,#], Jiayang Xu[1], Haiwa Wang[2], Qi Zhu[2], Wen Shi[6], Yue Fang[3], Fan Jiang[2], Shanbao Tong[1], Yunting Zhang[4,*], Xiaoli Guo[1,*]

[1] School of Biomedical Engineering, Shanghai Jiao Tong University, Shanghai, China.
[2] Department of Developmental and Behavioral Pediatrics, National Children's Medical Center, Shanghai Children's Medical Center, affiliated to School of Medicine Shanghai Jiao Tong University, Shanghai, China
[3] China Welfare Institute Nursery, Shanghai, China.
[4] Child Health Advocacy Institute, National Children's Medical Center, Shanghai Children's Medical Center, affiliated to School of Medicine Shanghai Jiao Tong University, Shanghai, China.
[5] Department of Computer Science, Vanderbilt University, Nashville, TN, USA.
[6] Department of Biomedical Engineering, Johns Hopkins University School of Medicine, Baltimore, MD, USA.

[#] These authors contributed equally to this work.
* Corresponding authors:
　　Dr. Xiaoli Guo at School of Biomedical Engineering, Shanghai Jiao Tong University, Shanghai, China, e-mail: meagle@sjtu.edu.cn.
　　Dr. Yunting Zhang at Child Health Advocacy Institute, National Children's Medical Center, Shanghai Children's Medical Center, affiliated to School of Medicine Shanghai Jiao Tong University, Shanghai, China, e-mail: edwinazhang@shsmu.edu.cn.

**Data availability statements:** Anonymised data is available from the authors upon request.
**Funding statement:** This work was supported by the Innovative Research Team of High-level Local Universities in Shanghai.
**Competing interests:** The authors declare that they have no competing interests.
**Ethics approval statement:** The study was approved by the Review Board of Ethics Committee of Shanghai Children's Medical Center (SCMCIRB-K2020108-1).

## Abstract

Mother-child interaction is highly dynamic and reciprocal. Switching roles in these back-and-forth interactions serves as a crucial feature of reciprocal behaviors while the underlying neural entrainment is still not well-studied. Here, we designed a role-controlled cooperative task with dual EEG recording to study how differently two brains interact when mothers and children hold different roles. When children were actors and mothers were observers, mother-child inter-brain synchrony emerged within the theta oscillations and the frontal lobe, which highly correlated with children's attachment to their mothers. When their roles were reversed, this synchrony was shifted to the alpha oscillations and the central area and associated with mothers' perception of their relationship with their children. The results suggested an observer-actor neural alignment within the actor's oscillations, which was modulated by the actor-toward-observer emotional bonding. Our findings contribute to the understanding of how inter-brain synchrony is established and dynamically changed during mother-child reciprocal interaction.

# 1. Introduction

Humans are fundamentally social. Social interaction allows individuals of the same species to communicate with each other and such interaction is a matter of survival (Frith and Frith, 2007). Through the very early experience of interactions, children can coordinate their performance toward the smooth execution of social goals and co-related social needs to maximize well-being and thrive (Endevelt-Shapira et al., 2021). Therefore, adult-child interaction, especially the interactions between mother and child, is essentially important for children's social development.

Recently, with the advent of hyperscanning techniques, researchers have been able to characterize the inter-brain synchrony by monitoring two or more brains concurrently (Czeszumski et al., 2020; Lindenberger et al., 2009; Montague et al., 2002; Wang et al., 2018), which sheds light on the neural mechanisms of interactions. By implementing this approach to the adult-child interaction studies, connections have been established between some interaction factors and inter-brain synchrony. For example, children always showed higher inter-brain synchrony with their parents than with a stranger (Endevelt-Shapira et al., 2021; Reindl et al., 2018; Zhao et al., 2021). The inter-brain synchrony between children and their parents was uniformly found to be higher in cooperative tasks in comparison to when they solved the same task individually (Nguyen et al., 2020; Reindl et al., 2018; Wang et al., 2020). Moreover, communication signals, such as maternal chemosignals (Endevelt-Shapira et al., 2021), positive emotion (Santamaria et al., 2020), affectionate touch (Nguyen et al., 2021a), as well as direct gaze from an adult stranger (Leong et al., 2017), would elicit significant neural entrainment in children. Given these pioneering studies, inter-brain synchrony has been considered a biomarker for mutual task engagement (Wass et al., 2020) and might help create an optimal learning environment for children (Atzil and Gendron, 2017; Nguyen et al., 2020). However, these studies only calculated the average level of inter-brain synchrony across the whole interaction (Endevelt-Shapira et al., 2021; Leong et al., 2017; Miller et al., 2019; Nguyen et al., 2021b, 2020; Quiñones-Camacho et al., 2020; Reindl et al., 2018; Santamaria et al., 2020; Wang et al., 2020) or simply targeted the unidirectional communication signals from adults towards children (Endevelt-Shapira et al., 2021; Leong et al., 2017; Nguyen et al., 2021a; Santamaria et al., 2020). Therefore, the evidence thus far is still insufficient to uncover the dynamic coupling between mothers and children during the back-and-forth interactions in daily life.

Mother-child interaction is by nature marked by the "serve and return" behaviors, termed mutual reciprocity (George, 1996). Dyadic reciprocal social interaction requires social partners to explicitly take on complementary and alternating roles, such as sender and receiver, throughout the course of interactions (Redcay and Schilbach, 2019). Recently, increasing studies have demonstrated that even in very early ages, young children acted not only as purely passive recipients of information, but also played an active role as senders (Murray and Trevarthen, 1986; Stahl and Feigenson, 2015; Wass et al., 2020). The roles that each partner takes on are usually not fixed, but alternate throughout the interactions (Schilbach et al., 2013; Wass et al., 2020). Hence, distinguishing roles when investigating the inter-brain synchrony during mother-child interaction can largely facilitate modeling and understanding of the mutual reciprocity and might be a key to revealing the neural mechanism underlying the various dynamic social-cognitive processes involved.

Within the existing studies, some findings from naturalistic experimental designs indicated that in mother-child interactions, mothers and children tended to have distinct ways of behavioral and neural alignment towards each other when holding different roles. A study in physiological entrainment found that mothers increased their arousal level to match their child's, with a greater arousal response in mothers associated with lower subsequent child arousal (Wass et al., 2019). Another EEG study observed similar patterns in the behavior-brain association. Theta band oscillations in infants were found to be associated with attentional and encoding processes, while in adults a similar functional relationship was observed but at a higher frequency in the alpha band (Wass et al., 2018). However, during mother-infant joint play, the frequency of the mother's highest association between EEG power and attention was down-shifted to the theta range—similar to the infant's peak frequency of association. During this process, when the mother was more neurally responsive, the infant was more attentive (Wass et al., 2018). This study suggested that EEG frequency shifting might be a unique process for mothers to neurally align with their children for effective interactions. However, it still remains unknown whether this neural attunement is mutual or unidirectional as well as its potential association with inter-brain alignment. Therefore, it is worth probing into the roles in mother-child interaction and investigating the dynamic properties of this neural entrainment.

Additionally, previous hyperscanning studies also suggested that the mother-child relationship might influence the neural entrainment. When mothers and children were viewing their pre-recorded interaction vignettes, brain-to-brain synchrony only emerged during the perception of synchronous patterns specific to the attachment context (Levy et al., 2017). State-like factors such as maternal stress (Nguyen et al., 2020) and child irritability (Quiñones-Camacho et al., 2020), were also found to be associated with neural synchronization during the task. Nevertheless, whether and how the mother-child relationship, as well as their interaction experience, influence their inter-brain synchrony in different role contexts is yet to be discovered.

To overcome the above limitations and examine the concurrent inter-brain entrainment along with the alternating roles during mother-child mutual-reciprocal interaction, we designed a role-controlled tangram-solving task which comprised two role contexts, where children acted as actors/senders (who execute the action) and observers/receivers (who observe the action), respectively. We evaluated the effect of interaction in each role context by comparing a high-interactive condition (cooperation) with a low-interactive (accompanied) and a non-interactive (individual) condition, and compared between mother-child and stranger-child interactions. By applying dual EEG recording, we strived to answer the following questions: (1) How do roles in mother-child interaction influence the neural entrainment between the two brains of dyads? (2) Is the inter-brain synchrony in two role contexts distinguishable between mother-child and stranger-child interactions? (3) Is the inter-brain synchrony in two role contexts associated with the relationship between mothers and their children?

## 2. Methods

### 2.1 Participants

Forty-two preschool children were recruited from China Welfare Institute Nursery, Shanghai. Twenty-one children participated with their biological mothers (mother-child dyads), and the others performed the task with an unfamiliar female adult experimenter (stranger-child dyads). All the participants were right-handed according to the Edinburgh handedness inventory, with normal or corrected-to-normal vision, and had no known neurological or psychiatric disorders.

Two mother-child dyads and two stranger-child dyads were excluded due to that the children did not well follow the experimental instructions. Another mother-child dyad was excluded because of poor EEG quality. The final sample thus consisted of 18 mother-child dyads (children: 10 boys and 8 girls, mean age: 3.83 ± 0.13 years) and 19 stranger-child dyads (children: 6 boys and 13 girls, mean age: 3.88 ± 0.11 years). There was no significant difference in children's sex ($\chi^2(1) = 2.165$, $P = 0.141$) and age ($t(35) = -0.314$, $P = 0.755$) between mother-child and stranger-child groups. Each subject (parents on behalf of their child) signed a written informed consent prior to participation. The study was approved by the Review Board of Ethics Committee of Shanghai Children's Medical Center (SCMCIRB-K2020108-1).

### 2.2 Behavioral assessments

Children's IQ was measured with the Chinese version of Wechsler Preschool and Primary Scale of Intelligence-Fourth Edition (WPPSI-IV) (Wechsler, 2012) which includes verbal and performance tests. The verbal test includes subtests of *Information*, *Vocabulary*, *Arithmetic*, *Similarities*, and *Comprehension*; and the performance test includes subtests of *Zoo Locations*, *Picture Completion*, *Object Assembly*, *Coding*, and *Block Design*. Among all subtests, *Object Assembly* (OA) measures skills similar to those required for tangram puzzle-solving. All subtests were then aggregated into a full-scale IQ (FSIQ). FSIQ score and OA subscore (IQ-OA) were used to measure children's cognitive development. Each child was assessed by two standardized trained assessors. One child in the mother-child group couldn't complete the FSIQ test but only the OA subtest. There was no significant group difference in children's FSIQ score (mother-child group: FSIQ = 111.65 ± 3.25, stranger-child group: FSIQ = 117.58 ± 2.76; $t(34) = -1.4$, $P = 0.171$) as well as their OA subscore (mother-child group: OA = 12.94 ± 1.40, stranger-child group: OA = 16.11 ± 1.39; $t(35) = -1.600$, $P = 0.117$).

Children's psychosocial well-being was assessed with the Chinese version of the teacher-reported Strengths and Difficulties Questionnaire (SDQ) (Goodman and Goodman, 2009). The questionnaire contains 25 items in five dimensions (emotional symptoms, conduct problems, hyperactivity or inattention, peer relationship problems, and prosocial behavior), and each item was scored as 0 = not true, 1 = somewhat true or 2 = certainly true. The total score in the first four dimensions (excluding prosocial behavior) was generated as a total difficulties score (TDS) indicative of psychosocial problems. Using the recommended cut-off values of TDS in China(Du et al., 2008), which had been previously validated for children aged 3-16 years, there was only one child in mother-child group categorized as 'borderline'. There was no significant difference in children's TDS ($t(35) = .536$, $P = 0.595$) between the mother-child (TDS = 7.33 ± 0.81) and stranger-child (TDS = 6.68 ± 0.89) groups.

In the mother-child group, mothers' parenting stress and mother-child attachment were also measured. Mothers' parenting stress was assessed with a 36-item self-report Parenting Stress Index-short Form (PSI-SF) (Abidin, 1995; Haskett et al., 2006). It includes three subscales, which are *Parental Distress* (PD), *Parent-Child Dysfunctional Interaction* (P-CDI), and *Difficult Child* (DC). The PD subscale indicates the level of distress resulting from personal factors such as depression or conflict with a partner and from life restrictions due to the demands of child-rearing. The P-CDI subscale provides an indication of parents' dissatisfaction with interactions with their children and the degree to which parents find their children unacceptable. The DC subscale measures parents' perceptions of their children's self-regulatory abilities. Each subscale consists of 12 items and each item was scored on a five-point scale from 5 (strongly agree) to 1 (strongly disagree). The higher the score, the more parenting stress.

The child attachment relationship scale (Ye, 2011) was self-reported by the mother. There are 21 items covering three dimensions at security attachment (SA), avoidant or disorganized attachment, and ambivalent attachment. Each item was scored from 1 (strongly disagree) to 4 (strongly agree) points. Scores of items in each dimension were added, respectively, and the sum was calculated as the value of each dimension. Children's parent-child attachment was assigned to the dimension with the highest value, and all children in the mother-child group in this study were classified as security attachment. Therefore, we only included the SA score in further analysis.

## 2.3 Experiment design

The adult and the child in each dyad were asked to participate in a tangram puzzle-solving video game on a touch screen (23.8-inch, NEWTAP Inc., China). The game was designed attractive and easy to accomplish for preschool children. Each tangram template consisted of seven pieces of geometric shape puzzles, and subjects were asked to move puzzles using their right hand.

There were twelve different tangram templates that were equally and randomly assigned for games in individual, accompanied, and cooperative conditions, respectively. In the individual condition, one participant solved all the puzzles in the template alone and the other one was seated in another room watching the screen recording from the actor. In the accompanied condition, one participant solved all puzzles alone while the other was sitting next to him/her and watching the screen. In these two conditions, each person solved two tangram templates in each condition. While in the cooperative condition, the two participants were seated next to each other and jointly solved four tangram templates together by moving the puzzle pieces by turn. Among different interactive conditions, the fundamental events in solving each piece of the puzzles were the same, i.e., one participant solving the puzzle (the actor) and the other observing partner's activity (the observer), but with different degrees of interaction. Each interactive condition thus comprised two role contexts, i.e., the adult was the actor while the child was the observer or their roles were reversed. Therefore, there were 14 movements × 3 interactive conditions × 2 role contexts for each participant. The participants were asked to move puzzles steadily and keep attending and silent during the whole procedure. The experiment design is illustrated in Fig. 1.

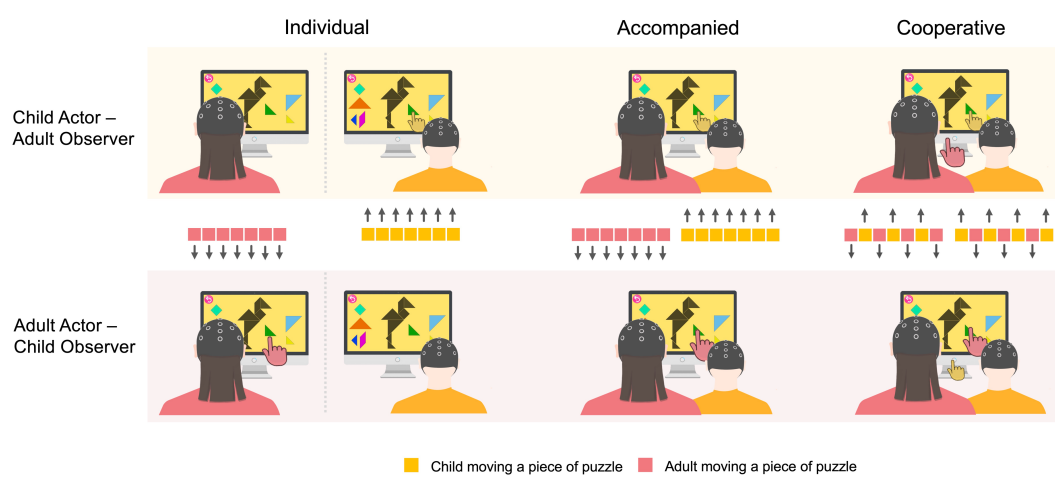

**Fig. 1. Illustration of experiment design.** Both the mother-child and stranger-child dyads were asked to solve tangram puzzles in individual, accompanied and cooperative conditions. In the individual condition, one participant solved all puzzles alone and the other was seated in another room watching the screen recording from the actor. In the accompanied condition, one participant solved all the puzzles alone while his/her partner was sitting next to him/her, watching the screen. In the cooperative condition, the participants were seated next to each other and jointly solved tangram templates by moving the puzzle pieces by turn. Although the degree of interaction was different in three interactive conditions, the fundamental events in solving each piece of the puzzles were the same and could be divided into two contexts according to the participants' roles. One was the child solving the puzzle (child as the actor) and the adult observing the child's activity (adult as the observer), and the other was with their roles reversed (adult as the actor and child as the observer).

## 2.4 Data acquisition

During the experiment, EEG signals from the child and the adult were recorded simultaneously using two 32-channel mobile EEG systems (NeuSen.W32, Neuracle, China) with electrodes placed according to the international 10/20 system. The synchronization of two EEG systems was achieved by sending a trigger signal simultaneously to both EEG systems. For each EEG system, the sample rate was set at 1000 Hz, with FCz as the reference electrode and AFz as the ground electrode. The electrode impedances were kept below 10 kΩ during the recording.

The whole procedure was also monitored and recorded by two cameras. One was set behind the participants to capture their finger actions on the screen, and the other was placed in front to monitor their attention and interactions. Video recordings were synchronized with the EEG systems.

**2.5 EEG preprocessing**

The EEG preprocessing was performed to obtain attentive and artifacts-free neural signals during each movement of a puzzle piece using the EEGLAB toolbox (Delorme and Makeig, 2004) in MATLAB (MATLAB R2018b, The MathWorks Inc., Natick, MA). Peripheral temporal channels (T7, T8, Tp9, Tp10) were excluded from the analysis due to the significant contamination of muscular artifacts, and therefore, a total of 26 channels (Fp1, Fp2, F3, F4, Fz, F7, F8, FC1, FC2, FC5, FC6, C3, C4, Cz, CP1, CP2, CP5, CP6, P3, P4, Pz, P7, P8, O1, O2, Oz) were selected for analysis. EEG signals were band-pass filtered into 1-35Hz and notch filtered at 50 Hz. Filtering was done on continuous data to avoid edge effects. Independent component analysis (ICA) was further utilized to remove ocular and muscular artifacts. Then, the EEG data were filtered into children's theta (3-6 Hz) and alpha (6-9 Hz) bands.

EEG data in theta and alpha bands were segmented according to each movement of a puzzle piece. The time window of interest (WOI) was selected from the start to the end times of moving the puzzle piece by manually labeling the video frames. As the duration varied across movements and participants, a 1-s epoch at the middle of each WOI was selected for further analysis. Those epochs when the participants were speaking or not attentive according to the video, or with amplitude values exceeding ±100μV were excluded. Finally, 12.85 ± 0.12 epochs were preserved on average from each participant in each interactive condition and each role context.

**2.6 Inter-brain synchrony and thresholding**

The inter-brain synchrony was characterized by phase locking value (PLV) which measures the consistency of the phase variation over a period of time between two band-pass filtered signals (Lachaux et al., 1999).

Briefly, the instantaneous phase sequences of two narrow-band signals $x(t)$ and $y(t)$, which were $\varphi_x(t)$ and $\varphi_y(t)$, were calculated using Hilbert transform. Sequentially, the PLV between two signals can be derived as

$$PLV_{x,y} = \frac{1}{N}\left|\sum_{t=1}^{N} exp^{j\left(\varphi_x(t)-\varphi_y(t)\right)}\right|, \quad (1)$$

where *N* represents the samples within the selected time window. For each dyad, each interactive condition, each context, and each frequency band, the PLV values between the child's channels and the adult's channels were calculated, forming a $26 \times 26$ asymmetric matrix.

As the common intrinsic properties of the EEG signals and shared external perturbation, like the same experimental environment and similar visual stimuli, could also cause synchrony between participants in each dyad (Burgess, 2013; Leong et al., 2017; Santamaria et al., 2020), a threshold method based on surrogating was implemented to reduce the spurious synchrony that was uncorrelated to the task. Specifically, a surrogate dataset (1000 permutations) was generated by shuffling the time frames between adult and child in each dyad and interactive condition, which means randomly pairing the adult's (child's) epochs to the child's (adult's) epochs at different time. PLV was calculated for each permutation and then averaged. Sequentially, for each dyad, a threshold matrix was derived by averaging the surrogate connectivity matrices across conditions in the same context. The inter-brain connections with their strength significantly exceeding their respective threshold values (one tailed paired *t*-test, $P < 0.05$) were regarded as the "real" connections and preserved for further analysis from two following aspects.

First, the global inter-brain network density (IBD) and strength (IBS) were calculated. The IBD was calculated as the ratio of real connections to all possible connections. The IBS was used to quantify the average intensity of inter-brain synchrony. To ensure the comparability of IBS among different interactive conditions, the least number of significant connections across all three interactive conditions was used for average in each context. Second, at the local level, the node degree and regional IBS were calculated to explore the brain regions involved in child-adult interaction. Node degree is the total number of edges connected to a certain node, and the regional IBS is the average intensity of inter-brain synchrony between specific brain regions.

**2.7 Statistical analysis**

Two-way repeated measures ANOVAs with Greenhouse-Geisser correction for sphericity violation were performed on the IBS (both global and regional IBS) in each role context (adult-actor/child-observer and child-actor/adult-observer), with a within-group factor of *interactive condition* (3 levels: individual, accompanied and cooperative conditions) and a between-group factor of *group* (2 levels: mother-child dyads and stranger-child dyads). Bonferroni correction was used to adjust for multiple comparisons in post-hoc analysis. Pearson's correlation (for the variables conforming to the normal distribution) or Spearman's correlation (for the variables not conforming to the normal distribution) was applied to explore the relationship between inter-brain synchrony and behavioral measures (including SA, PD, P-CDI, DC, and TDS scores). The normality of variables was tested by Shapiro-Wilk's test. Statistical significance was accepted for $P < 0.05$. False discovery rate (FDR) was used for multiple comparisons correction in regional IBS and correlation analysis. All data are presented as mean ± standard error of the mean. All the statistical analyses were conducted using IBM SPSS Statistics 24.0 and MATLAB R2018b.

## 3. Results

### 3.1 Inter-brain network between child-actor and adult-observer

*3.1.1 Global network metrics*

When the child was the actor solving puzzles and his/her adult partner was the observer, their inter-brain network was found to be modulated by interaction mainly in the theta band (Fig. 2). The theta-band inter-brain network was more densely connected with increasing degree of interaction (Fig. 2A and 2C), in either the mother-child (individual: theta-IBD = 0.032, accompanied: theta-IBD = 0.037, cooperative: theta-IBD = 0.109) or the stranger-child (individual: theta-IBD = 0.018, accompanied: theta-IBD = 0.018, cooperative: theta-IBD = 0.043) dyads. However, the theta-IBD in each interactive condition as well as its increment with interaction degree was larger and more prominent in the mother-child dyads compared with the stranger-child dyads.

Likewise, the strength of theta-band inter-brain synchrony also showed a similar modulation by interaction degree (Fig. 2D). Two-way repeated measures ANOVA on the theta-IBS revealed a significant main effect of *interactive condition* ($F(2,70) = 22.611, P < 0.001, \eta^2_p = 0.392$), exhibiting a stronger theta-band inter-brain synchrony in cooperation than accompanied and individual conditions (individual: theta-IBS = 0.412 ± 0.003, accompanied: theta-IBS = 0.412 ± 0.004, cooperative: theta-IBS = 0.439 ± 0.003). Post-hoc analysis found that the theta-IBS in cooperation was significantly stronger than individual ($P_{Bonf} < 0.001$) and accompanied ($P_{Bonf} = 0.006$) conditions, while no significant difference was found between individual and accompanied conditions ($P_{Bonf} = 1.000$). The main effect of *group* was also significant ($F(1,35) = 40.864, P < 0.001, \eta^2_p = 0.539$). The mother-child dyads displayed a stronger theta-IBS (theta-IBS = 0.433 ± 0.003) than the stranger-child dyads (theta-IBS = 0.409 ± 0.003) when the child was the actor.

In the alpha band (Fig. 2B, E, and F), the main effect of *interactive condition* was also significant ($F(2,70) = 11.813, P < 0.001, \eta^2_p = 0.252$). However, contrary to the increasing trend of inter-brain synchrony in the theta-band, the strength of inter-brain synchrony in the alpha band displayed a declining trend (individual: alpha-IBS = 0.436 ± 0.004, accompanied: alpha-IBS = 0.425 ± 0.003, cooperative: alpha-IBS = 0.414 ± 0.003, Fig. 2F).

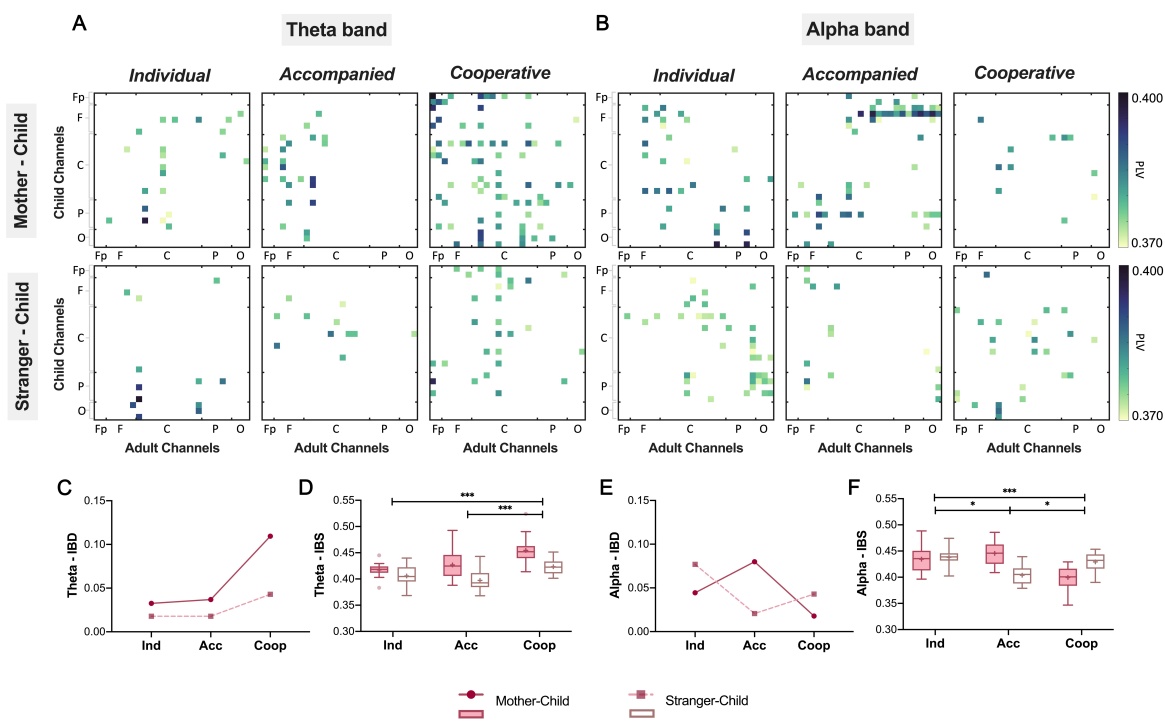

Fig. 2. Global features of inter-brain network of mother-child and stranger-child dyads when the adult was observing the child solving puzzles in individual, accompanied, and cooperative conditions. (A) and (B) show the inter-brain connectivity matrices in the theta (A) and alpha (B) bands, respectively. Rows of matrices denote the child's channels, and columns denote the adult's channels. In each matrix, only connections with their strength significantly exceeding the corresponding threshold obtained from surrogating (one tailed paired $t$-test, $P < 0.05$) are shown as small squares, and the color of squares stands for their PLV strength. (C) shows inter-brain density in the theta band (theta-IBD), and (D) shows inter-brain strength in the theta band (theta-IBS). (E) shows inter-brain density in the alpha band (alpha-IBD), and (F) shows inter-brain strength in the alpha band (alpha-IBS). * $P_{Bonf} < 0.05$, *** $P_{Bonf} < 0.001$.

### 3.1.2 Local network metrics

Since the increase of inter-brain synchrony with interaction degree was prominent in the theta band especially when the mother was observing her child solving puzzles, we further calculated the node degree and regional IBS of mother-child inter-brain network in the theta band to explore important brain regions involved in mother(observer)-child(actor) interaction. Distribution of node degree suggested that there were more theta-band inter-brain connections between child's left (pre)frontal (Fp1, F7), right fronto-centro-parietal (FC2, FC6, CP2, Pz) and temporoparietal (CP6, P8) channels, and mother's (pre)frontal (Fp1, Fp2, F3), right fronto-centro-parietal (FC2, Cz, CP2) and left temporospatial (CP5) channels in cooperative condition (each node degree ≥ 4, Fig. 3A). Moreover, regional theta-IBS (Fig. 3B) showed overwhelmingly strongest inter-brain synchrony in cooperative condition between almost all regions of two brains.

However, a significant increase with interaction degree (main effect of *interactive condition*, $P_{FDR}$ < 0.05) was only found between child's (pre)frontal and mother's prefrontal areas (Child_Fp – Mother_Fp and Child_F – Mother_Fp) and between child's central and mother's frontal areas (Child_C – Mother_F).

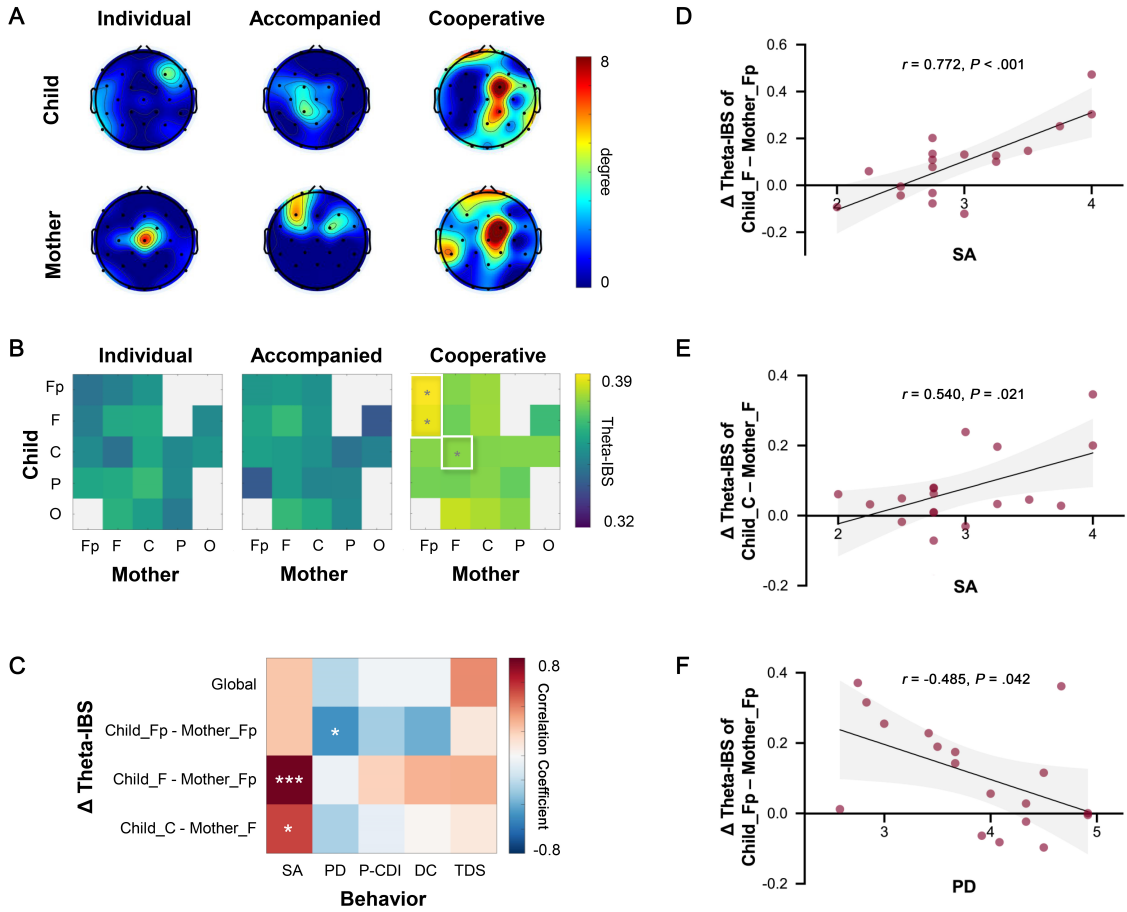

Fig. 3. Local features of mother-child inter-brain network in the theta band when the mother was observing her child solving puzzles in individual, accompanied and cooperative conditions. (**A**) shows topographies of node degree of inter-brain networks. (**B**) shows theta-band inter-brain strength between specific brain regions (i.e., regional theta-IBS), including prefrontal (Fp), frontal (F), central (C), parietal (P), and occipital (O) regions. * $P_{FDR}$ < 0.05 for the main effect of *interactive condition*. (**C**) Correlation between behavior measures and the increment of theta-IBS from accompanied to cooperative condition (Δtheta-IBS), and significant correlations are shown with scatter plots (**D-F**). SA: security attachment score, PD: parental distress score, P-CDI: parent-child dysfunctional interaction score, DC: difficult child, TDS: total difficulties score. *$P$ < 0.05, ***$P$ < 0.001.

### 3.1.3 Correlations between theta-band mother-child IBS and behavior

The global theta-IBS as well as the regional theta-IBS of Child_Fp – Mother_Fp, Child_F – Mother_Fp and Child_C – Mother_F connectivity, which were significantly enhanced with interactive degree when the mother was observing her child solving puzzles, were correlated with behavioral measures (Fig. 3C). The increment of theta-IBS from accompanied to cooperative condition was found significantly and positively correlated with the SA score at Child_F – Mother_Fp ($r(18) = 0.772$, 95%CI: [0.477, 0.911], $P < 0.001$, Fig. 3D) and Child_C – Mother_F ($r(18) = 0.540$, 95%CI: [0.098, 0.804], $P = 0.021$, Fig. 3E). Furthermore, the increment of Child_Fp – Mother_Fp theta-IBS also showed a significant negative correlation with the PD score in PSI-SF ($r(18) = -0.485$, 95%CI: [-0.776, -0.023], $P = 0.042$, Fig. 3F). However, no significant correlation was found between the global theta-IBS and behavioral measures. Note that the correlation between the increment of Child_F – Mother_Fp theta-IBS and the SA score survived after FDR correction ($P_{FDR} = 0.003$).

**3.2 Inter-brain network between adult-actor and child-observer**

*3.2.1 Global network metrics*

When the child was observing the adult partner solving puzzles, interestingly, interaction-related modulation on their inter-brain network was primarily shown in the alpha band (Fig. 4). IBD in the alpha band was increased with increasing degree of interaction (Fig. 4B and 4E), especially in the mother-child dyads (individual: alpha-IBD = 0.012, accompanied: alpha-IBD = 0.036, cooperative: alpha-IBD = 0.112). Similarly, the stranger-child dyads also showed a slight increase in the alpha-IBD with increasing degree of interaction (individual: alpha-IBD = 0.012, accompanied: alpha-IBD = 0.047, cooperative: alpha-IBD = 0.067).

The alpha-IBS also showed a significant enhancement with increasing degree of interaction (individual: alpha-IBS = 0.376 ± 0.004, accompanied: alpha-IBS = 0.427 ± 0.005, cooperative: alpha-IBS = 0.446 ± 0.004, Fig. 4F), and this enhancement was statistically demonstrated by a significant main effect of *interactive condition* ($F(2,70) = 64.764$, $P < 0.001$, $\eta^2_p = 0.649$). Post-hoc analysis found that the alpha-IBS in accompanied was significantly stronger than individual condition ($P_{Bonf} < 0.001$), and the alpha-IBS in cooperation was further enhanced compared with accompanied ($P_{Bonf} = 0.028$) and individual ($P_{Bonf} < 0.001$) conditions. Additionally, we also observed a significant main effect of *group* ($F(1,35) = 4.315$, $P = 0.045$, $\eta^2_p = 0.110$), exhibiting stronger alpha-IBS in the mother-child dyads (alpha-IBS = 0.422 ± 0.004) than the stranger-child dyads (alpha-IBS = 0.411 ± 0.004).

In the theta band, only the stranger-child dyads showed an increasing trend of inter-brain synchrony when the child was the observer (Fig. 4A, C, and D). The stranger-child dyads showed denser and stronger theta-band inter-brain connectivity with increasing degree of interaction (individual: theta-IBD = 0.015, theta-IBS = 0.400 ± 0.005; accompanied: theta-IBD = 0.033, theta-IBS = 0.426 ± 0.005; cooperative: theta-IBD = 0.058, theta-IBS = 0.440 ± 0.005). However, in the mother-child dyads, the theta-IBD and IBS were lower in accompanied condition compared with individual and cooperative conditions (individual: theta-IBD = 0.037, theta-IBS = 0.433 ± 0.006; accompanied: theta-IBD = 0.018, theta-IBS = 0.405 ± 0.004; cooperative: theta-IBD = 0.025, theta-IBS = 0.419 ± 0.007).

**Fig. 4. Global features of inter-brain network of mother-child and stranger-child dyads when the child was observing the adult solving puzzles in individual, accompanied, and cooperative conditions.** (**A**) and (**B**) show the inter-brain connectivity matrices in the theta (**A**) and alpha (**B**) bands, respectively. Rows of matrices denote the child's channels, and columns denote the adult's channels. In each matrix, only connections with their strength significantly exceeding the corresponding threshold obtained from surrogating (one tailed paired *t*-test, $P < 0.05$) are shown as small squares, and the color of squares stands for their PLV strength. (**C**) shows inter-brain density in the theta band (theta-IBD), and (**D**) shows inter-brain strength in the theta band (theta-IBS). (**E**) shows inter-brain density in the alpha band (alpha-IBD), and (**F**) shows inter-brain strength in the alpha band (alpha-IBS). * $P_{Bonf} < 0.05$, *** $P_{Bonf} < 0.001$.

*3.2.2 Local network metrics*

In order to explore the important brain regions involved in the mother(actor)-child(observer) interaction, we further analyzed the node degree and regional IBS of alpha-band inter-brain network. Distribution of node degree suggested that there were more alpha-band inter-brain connections between the child's right (pre)frontal (Fp2, F4, FC6) and left centro-parietal (CP1, CP5, P3, P7) channels, and the mother's left fronto-central (Fp1, F3, F7, FC5, C3) and right centro-parietal (Cz, CP2, P4) channels in cooperative condition (each node degree ≥ 4, Fig. 5A). Regional IBS (Fig. 5B) showed overwhelmingly strongest alpha-band inter-brain synchrony in cooperative condition, with significant increase with interaction degree (main effect of *interactive condition*, $P_{FDR} < 0.05$) between child's centro-parietal and mother's central areas (Child_C – Mother_C and Child_P – Mother_C) and between child's central and mother's prefrontal areas (Child_C – Mother_Fp).

### 3.2.3 Correlations between alpha-band mother-child IBS and behavior

The global alpha-IBS as well as the regional alpha-IBS of Child_C – Mother_C, Child_P – Mother_C and Child_C – Mother_Fp connectivity, which were significantly enhanced with interactive degree when the child was observing his/her mother solving puzzles, were correlated with behavioral measures (Fig. 5C). No significant correlation was observed between the global alpha-IBS and behavioral measures. However, the increment of Child_C – Mother_C alpha-IBS from accompanied to cooperative condition was found significantly and negatively correlated with the P-CDI score in PSI-SF ($\rho(18) = -0.520$, 95%CI: [0.037, 0.806], $P = 0.027$, Fig. 5D).

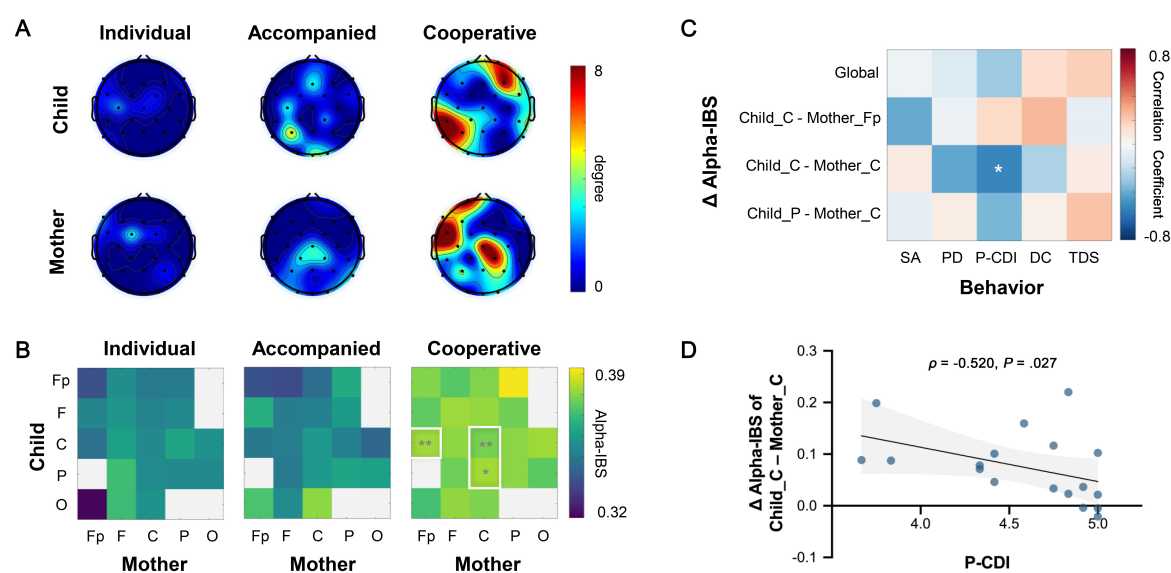

**Fig. 5. Local features of mother-child inter-brain network in the alpha band when the child was observing his/her mother solving puzzles in individual, accompanied, and cooperative conditions.** (**A**) shows topographies of node degree of alpha-band inter-brain networks. (**B**) shows alpha-band inter-brain strength between specific brain regions (i.e., regional alpha-IBS), including prefrontal (Fp), frontal (F), central (C), parietal (P), and occipital (O) regions. \*\**P_{FDR}* < 0.01, \**P_{FDR}* < 0.05 for the main effect of *interactive condition*. (**C**) Correlations between behavior measures and the increment of alpha-IBS from accompanied to cooperative condition (Δalpha-IBS), and the significant correlation is shown with scatter plot (**D**). SA: security attachment score, PD: parental distress score, P-CDI: parent-child dysfunctional interaction score, DC: difficult child, TDS: total difficulties score. \**P* < 0.05.

# 4. Discussion

Using a tangram puzzle-solving video game, we designed a well-controlled adult-child interaction experiment with gradually increasing interaction degree from individual, accompanied to cooperative conditions, and distinguished different interaction contexts where the child in the dyads acted as an actor/sender or an observer/receiver. By taking the advantage of high temporal resolution of EEG, we depicted the dynamic properties of inter-brain synchrony along with the alternating roles during adult-child interaction. Furthermore, by comparing the mother-child and stranger-child dyads as well as correlating the scores for mother-child relationship, we examined the association between inter-brain synchrony and inter-personal relationship.

Firstly, we found enhanced inter-brain synchrony during adult-child cooperation and demonstrated that the enhancement was actually induced by inter-personal interaction rather than shared external perturbation (e.g., common stimulation). Although previous studies have consistently reported increased inter-brain connectivity during adult-child interaction (Miller et al., 2019; Nguyen et al., 2021b, 2020; Reindl et al., 2018), the naturalistic experimental settings in these studies introduced more common stimulation to the cooperative than the individual condition, and therefore cannot exclude the spurious synchrony which was boosted by triggering similar neural responses to the common stimulation in both partners of the dyad (comment by Noreika *et al.* (Noreika et al., 2020)). In this study, we designed an individual condition with shared visual stimuli (i.e., one watching the screen recording of his/her partner solving the puzzle) as a control to reduce the spurious synchrony. Meanwhile, we also designed an accompanied condition between the individual and cooperative conditions. With such a careful experimental design, we found a gradual increase of inter-brain synchrony with the increasing degree of interaction (individual < accompanied < cooperative) in specific EEG frequency bands. Our results not only demonstrated the association between inter-brain synchrony and inter-personal interaction, but also revealed the importance of interaction quality or engagement in adult-child interaction with objective neural electrophysiological evidence.

Most importantly, we captured the characteristics of inter-brain synchrony in response to the alternating roles during adult-child interaction, and for the first time, proposed an association between the roles in adult-child interaction and the EEG oscillations of inter-brain synchrony. Specifically, when the child was the actor and the adult partner was the observer, the interaction-induced inter-brain synchrony was found in the theta band; and when their roles were reversed, this synchrony was shifted to the alpha band. Noted that both theta and alpha rhythms have been suggested to be closely related to adult-child interaction by previous EEG hyperscanning studies (Endevelt-Shapira et al., 2021; Leong et al., 2017; Santamaria et al., 2020). Leong *et al.* found that direct gaze increased bidirectional inter-brain connectivity between the female experimenter and infants in both theta and alpha bands (Leong et al., 2017). Santamaria *et al.* found that mothers' emotion would influence the mother-infant inter-brain network in the alpha band (Santamaria et al., 2020). Endevelt-Shapira *et al.* found that maternal chemosignals enhanced adult-infant inter-brain synchrony specifically in the theta band (Endevelt-Shapira et al., 2021). Although these pioneering studies demonstrated the involvement of theta and/or alpha rhythms in adult-child interactions especially in mother-child interactions, the underlying mechanism of why and how these two rhythms synchronize across two brains is still insufficiently understood. In this study, through a turn-taking cooperation experimental design, we were able to distinguish different interaction contexts where dyads played different roles, and found the frequency band where the synchronization took place was highly dependent on the roles in adult-child interaction. The theta rhythm is vital for infants and preschool children in cognitive functioning, anticipatory and sustained attention, and encoding (Orekhova et al., 2006; Saby and Marshall, 2012; Sperdin et al., 2018; Wass et al., 2018). However, when the children grow into adults, this attention- and cognition-related oscillations were shifted to the alpha band (Marshall et al., 2002; Wass et al., 2018). Interestingly, we found that the inter-brain synchrony during adult-child interaction prominently arose in the frequency band related to the actor's attention and cognitive functioning. One possible mechanism here we speculated is that in real interaction, especially in high-engaged interaction like cooperation, the observer may shift his/her attention- and cognition-related EEG frequency to the actor's to achieve neural synchrony with the actor. The proposed EEG frequency shift during adult-child interaction has been observed by Wass and colleagues (Wass et al., 2018). That is, during mother-infant joint play, the attention-related EEG frequency of mother was down-shifted from her own alpha band to the infant's attention-related theta range (Wass et al., 2018), implying a mother-toward-child neural alignment. However, our results suggested that the neural alignment was not only from adults to children, but also vice versa, i.e., "from the observer to the actor". A mutual prediction theory can be used to explain the inter-brain synchrony during interaction (Kingsbury et al., 2019). During turn-taking cooperation, the observer tried to understand and predict the behavior of the actor, and to achieve this, the observer might implicitly shift his/her EEG frequency to the actor's and synchronized his/her brain activity with the actor's. We think the EEG frequency shift and synchronization from the observer to the actor (or from receiver to sender) should be a vital neuroelectrophysiological mechanism underlying adult-child interaction.

It should be noted that both the mother-child and stranger-child dyads presented the inter-brain synchrony within the actor's oscillations, implying that they shared some common neural mechanisms in interaction. However, the strength of inter-brain synchrony and its increment with interaction degree were more prominent in the mother-child dyads compared with stranger-child dyads. This is in line with previous studies which compared the average inter-brain synchrony between parent-child and stranger-child interactions. Endevelt-Shapira *et al.* observed greater inter-brain synchrony in parent-child dyads than in stranger-child dyads during free play (Endevelt-Shapira et al., 2021), while Reindl *et al.* found significant enhancement of inter-brain synchrony during cooperation only between parents and their children but not between strangers and children (Reindl et al., 2018). These findings demonstrated that the inter-brain synchrony can effectively reflect the relationship of individuals during interactions, and more than that, from our results, a possible neural mechanism can be inferred, i.e., it might be easier for the dyads with stronger emotional bonds like mothers and children to achieve neural alignment by shifting one's own EEG rhythms to the other's frequency.

The importance of emotional bonds in adult-child inter-brain synchrony was further supported by the significant correlation between the strength enhancement of inter-brain synchrony and parent-child relationship in the mother-child dyads, which also showed role-dependent characteristics. When the child was the actor and the mother was the observer, their theta-band inter-brain synchrony enhanced in cooperation was primarily between the child's frontal, central, and mother's (pre)frontal areas, and this increment was significantly and strongly correlated with the child's attachment to their mother (i.e., SA score). From the age of 12 months, children become aware of how to actively coordinate their attention to both the objects and the adults simultaneously and use gestures to affect the adults' behavior (Chris and Philip J., 1995). Children with secure attachment have been found to show more joint attention (Naber et al., 2007) or seek to share experiences with the caregiver more than children with other attachment types (Humber and Moss, 2005). In the current study, although all the children in the mother-child dyads were classified as security attachment, children with stronger secure attachment level to their mothers may be better at 'inviting' their mothers to interact with them, which potentially, in turn, enabled their mothers to be more engaged in interaction and easier to align with children's neural activity. In this process, the frontal areas of both sides of the dyads seemed to play a crucial role. As one of the vital brain regions associated with the mentalizing process, the frontal area has been implicated to be responsible for understanding and predicting others' intentions and joint attention (Pan et al., 2017; Schilbach et al., 2013; Wang et al., 2018). Both mothers' and children's frontal activities have shown modulations by mother-child attachment in accumulated single-brain neuroimaging studies (Feldman, 2017; Laurent and Ablow, 2012; Minagawa-Kawai et al., 2009; Perry et al., 2017; Young et al., 2017); however, the evidence from the inter-brain network is still lacking. Though Miller *et al.* had tried to explore the link between attachment and the parent-child inter-brain network (Miller et al., 2019), they only found a tendency that the avoidant child attachment might associate with less cooperation-related synchrony in the right frontopolar prefrontal cortex. To our knowledge, our work is the first to establish the correlation between attachment level and inter-brain synchrony, and associate this correlation with the frontal lobe. Herein, we conjectured that stronger attachment between mothers and children might facilitate the mentalizing process and contribute to their inter-brain synchrony in the frontal lobe.

When the mother was the actor, the central area was more prominent in mother-child cooperation. The alpha-band inter-brain synchrony between the mother's and child's central area displayed not only a significant increase with interaction degree, but also a negative correlation with the mother's perception of their relationship with their child

(i.e., P-CDI score). It means, the more dissatisfaction that the mother felt when interacting with her child in daily life, the less increment of central-central inter-brain synchrony in cooperation. One recent study by Leong *et al.* also highlighted the role of central region during mother-child interaction, where they observed that the concurrent alpha-band neural entrainment across the central electrodes of mother-infant dyads was predictive of infants' social learning (Leong et al., 2019). Parent-child interactions were considered to provide the primary social learning context from infancy (Smelser and Baltes, 2001), during which children's social learning is mainly accomplished through observing their parents' activities. Therefore, one possible interpretation for our results is that when children were observing their mothers solving puzzles, those who had better parent-child interactions during daily life reported by their mothers may be more advanced in their social learning abilities, thus found it easier to attend to and follow their mothers' intentions in cooperation. This was neurally indicated by higher increment of alpha-band inter-brain synchrony with their mothers in the central area.

It is worth mentioning that during mother-child cooperation, actor-observer inter-brain synchrony was modulated specifically by the emotional bonding from the actor toward the observer rather than the observer's affection toward the actor. These findings suggest that stronger actor-toward-observer emotional bonding in a mother-child relationship seems to have provided a better context for mutual interaction with higher increment of inter-brain synchrony.

Several limitations of this study should be mentioned. First, the relationship between inter-brain synchrony and behavioral traits was only explored with mother-child attachment and interaction quality in daily life but not with the interaction behavior during the task, since participants were asked to minimize unnecessary interaction behavior (such as talking and task-irrelevant interactive motions) to reduce EEG artifacts. In addition, mother-child attachment and interaction quality in daily life were assessed using parent-report questionnaires rather than direct observation, which may generate bias. However, the bias was minimized by choosing standardized and widely used instruments. Finally, the tangram puzzle-solving game was easy for most of the 3-year-old children, and therefore we were not able to detect any association between the task performances of children (e.g., willingness, speed, or accuracy) and the inter-brain synchrony. Further studies are strongly encouraged to explore the correlation between concurrent behaviors and inter-brain synchrony during interaction.

Overall, our study is the first in this area to reveal the dynamic properties of mother-child inter-brain synchrony along with the alternating roles during interaction. The role-specific effects were found on the inter-brain synchrony within different EEG oscillations and between different brain regions, which also showed a significant modulation by actor-toward-observer emotional bonding. Our results help us understand how inter-brain synchrony is established, maintained, and changed during this mutual-reciprocal interaction, and provide new insights into the neural mechanisms underlying social-cognitive processes in mother-child interactions, such as mentalization and social learning. Our findings also suggest that the interaction dynamics should be considered in future research on the inter-brain synchrony between adults and children, not only through computational approaches (Li et al., 2021), but also by specifying and considering vital elements (such as alternating roles) of the interaction dynamics in experimental design.

## Acknowledgments

This work was supported by the Innovative Research Team of High-level Local Universities in Shanghai. We thank Dr. Xiaoning Sun and Dr. Jin Zhao in Shanghai Children's Medical Center for their valuable advice on assessment of parent-child interaction and child development. We also thank Dan Wang, Jiaqi Wang, Jiawen Wu and Ziping Xing for the help in data collection.

**Competing interests:** The authors declare that they have no competing interests.